\documentclass[11pt,a4paper,onecolumn]{article}

\usepackage{epsfig}
\usepackage{url}
\usepackage{amssymb,amsmath}
\usepackage{color}

\begin{document}

\title{Neutron skin and centrality classification in high-energy heavy-ion collisions at the LHC}

\author{Hannu Paukkunen\footnote{hannu.paukkunen@jyu.fi} \\ \\
{\small Department of Physics, University of Jyv\"askyl\"a, P.O. Box 35, FI-40014 University of Jyv\"askyl\"a, Finland} \\ \\
{\small Helsinki Institute of Physics, P.O. Box 64, FI-00014 University of Helsinki, Finland}
}

\maketitle

\begin{abstract}

The concept of centrality in high-energy nuclear collisions has recently
become a subject of an active debate. In particular, the experimental
methods to determine the centrality that have given reasonable results for many observables in 
high-energy lead-lead collisions at the LHC have led to surprising behaviour in the 
case of proton-lead collisions. In this letter, we discuss the possibility to
calibrate the experimental determination of centrality by asymmetries
caused by mutually different spatial distributions of protons and neutrons inside the nuclei --- a 
well-known phenomenon in nuclear physics known as the neutron-skin effect.

\end{abstract}

\section{Introduction}

In high-energy heavy-ion experiments \cite{HighpTPhysicsintheHeavyIonEra,PhenomenologyofUltraRelativisticHeavyIonCollisions}
like those now performed at the Large Hadron Collider (LHC),
the collisions are often
categorized according to their centrality aiming to separate the central head-on collisions
from the peripheral ones in which only the edges of the nuclei collide. This has
been generally realized by sorting the events according to the amount
of particles or energy deposited in specific parts of the detector, details varying from one experiment
to another \cite{Abelev:2013qoq,Adam:2014qja,Chatrchyan:2011sx,Aad:2014eoa}.
In its simplicity, the idea is that in increasingly central collisions the colliding nuclei disintegrate
more completely thereby producing more particles.
In nucleus-nucleus collisions, an existence of a correspondence between the intuitive geometrical
notion of centrality and its experimental determination is supported by the systematics 
of azimuthal anisotropies \cite{Ollitrault:1992bk,Alver:2010gr} in the spectra of low-transverse-momentum (low-$p_T$) particles
\cite{Aad:2014eoa,Chatrchyan:2012ta,Aamodt:2010pa,Abelev:2012di} which,
in models based on classical collective dynamics \cite{Huovinen:2006jp,Gale:2013da,Huovinen:2013wma} 
are readily interpreted as reflecting the initial geometry of the collision zone.

Similar experimental procedures in proton-nucleus collisions at the LHC have revealed
much stronger centrality dependence of hard-process observables, like high-$p_T$ jets
\cite{Chatrchyan:2014hqa,ATLAS:2014cpa}, than what was anticipated e.g. from models for impact-parameter
dependent nuclear effects in parton distribution functions (PDFs) \cite{Helenius:2012wd,Helenius:2013bya}.
At the same time, however, the minimum-bias versions of the same jet observables --- obtained by ``integrating out''
the variable used for the centrality classification --- are in good agreement \cite{Chatrchyan:2014hqa,ATLAS:2014cpa,Paukkunen:2014pha} 
with the predictions based on collinearly factorized Quantum Chromo Dynamics (QCD).
This appears to indicate that the current experimental methods to realize the centrality
classification seriously interfere \cite{Tarafdar:2014oua,Adare:2013nff} with
the hard processes and, among other proposals \cite{Martinez-Garcia:2014ada,Bzdak:2014rca,Alvioli:2014eda,Perepelitsa:2014yta},
it has been argued \cite{Armesto:2015kwa} that even the standard energy-momentum conservation 
plays a significant role. On top of this, the presence of event-by-event fluctuations in the
initial profile of nucleons inside nucleus can further distort the multiplicity-based experimental
centrality classification. In the aggregate, the way that the centrality-selected measurements in proton-nucleus collisions should
be interpreted has turned out largely ambiguous. 
In nucleus-nucleus collisions interferences between the hard processes and the centrality
categorization are of less importance as the multiplicity of low-$p_T$ particles used in the events' centrality classification
is much larger and the correlations get diluted.

In this letter, we will discuss a centrality-dependent effect (in its geometric meaning) 
which stems from the fact that in spherical, neutron-rich nuclei the concentration of
neutrons is known to increase towards the nuclear surface \cite{Tarbert:2013jze,Abrahamyan:2012gp}.
We demonstrate how this causes effects in electroweak processes
that should be large enough to be measured at the LHC and could thereby
help in resolving open issues concerning the relationship between theoretical concept
of centrality and its experimental counterpart.

\section{Collision Geometry and Glauber Modeling}

\begin{figure}[ht!]
\begin{minipage}[b]{1.00\linewidth}
\center
\includegraphics[width=0.51\textwidth]{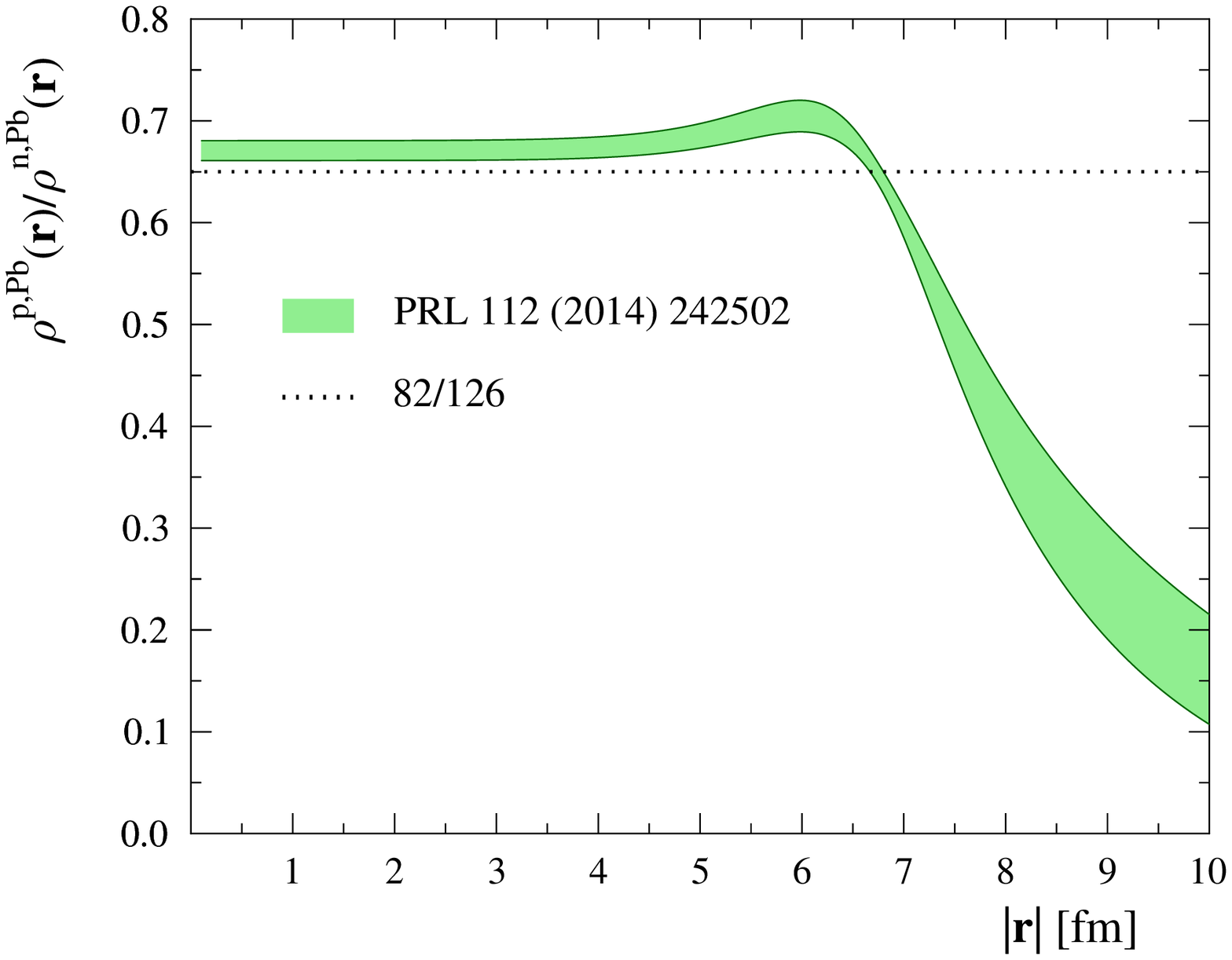}
\hspace{-0.9cm}
\includegraphics[width=0.51\textwidth]{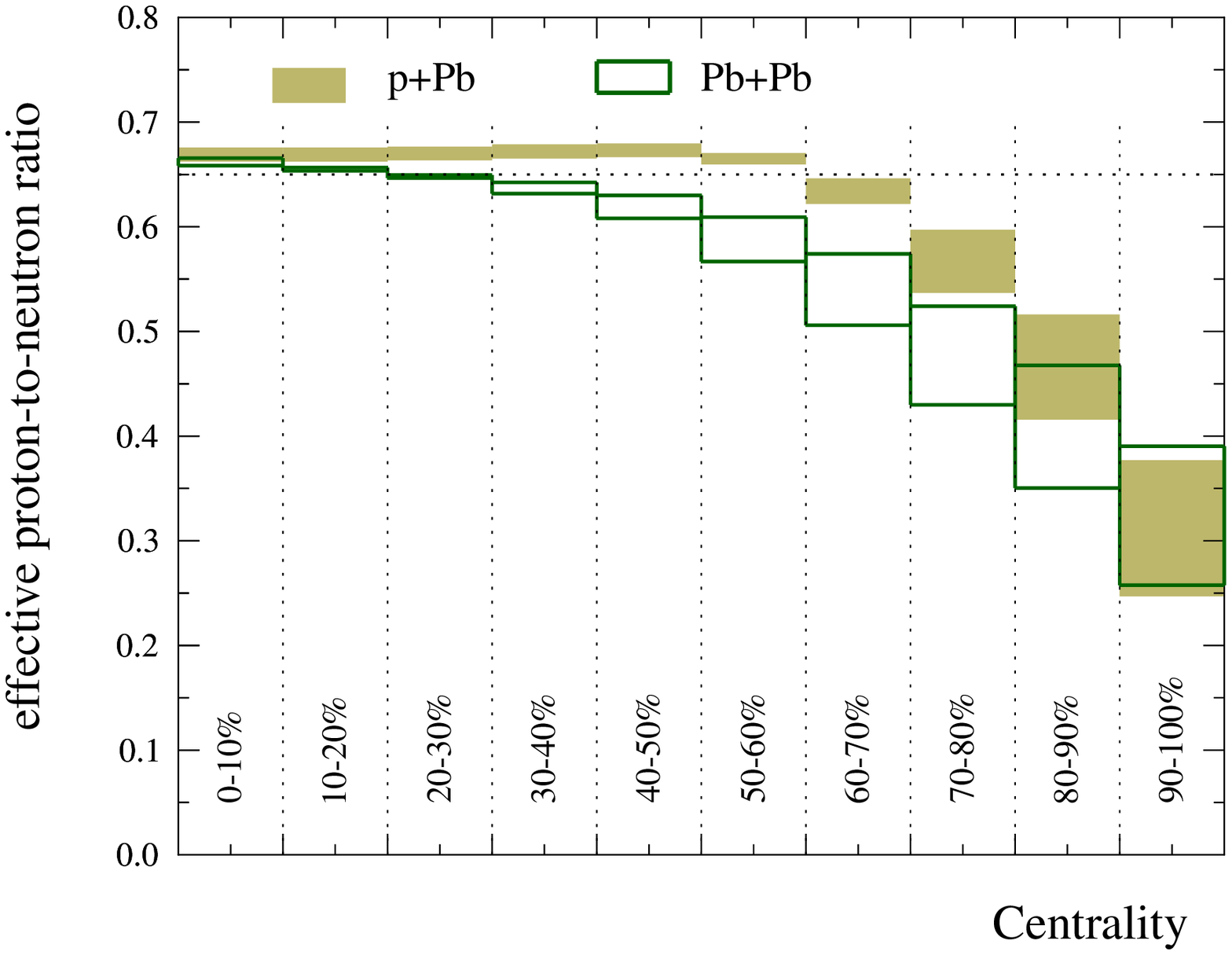}
\end{minipage}
\caption{Left-hand panel: The measured ratio of proton and neutron densities in $^{208}$Pb
as a function of nuclear radius. Right-hand panel: 
The ratio $Z_{\rm eff}({\mathcal{C}_k})/N_{\rm eff}({\mathcal{C}_k})$ for different
centrality classes in p+Pb (filled yellow rectangles) and Pb+Pb collisions (open green rectangles).
The heights of the rectangles are determined by the uncertainties given for the neutron density
in Ref.~\cite{Tarbert:2013jze}.
}
\label{Fig:densityratio}
\end{figure}

The density of nucleons $i$ in a spherical nucleus $A$ is often
parametrized using the two-parameter Fermi (2pF) distribution as
\begin{equation}
 \rho^{i,A}({\bf r}) = {\rho^{i,A}_0}/(1+e^{\frac{|{\bf r}|-d_i}{a_i}}),
\end{equation}
where the half-density parameter $d_i$ controls size of the nuclear
core and $a_i$ the thickness of the nuclear surface. The saturation densities
$\rho^{i,A}_0$ are determined by requiring the total amount of protons and neutrons
to remain constant (we consider only $^{208}$Pb nucleus in this paper),
\begin{equation}
\int d^3{\bf r} \rho^{{\rm p},A}({\bf r}) = Z = 82, \qquad \int d^3{\bf r} \rho^{{\rm n},A}({\bf r}) = N = 126.
\end{equation}
It is an experimental fact \cite{Tarbert:2013jze,Abrahamyan:2012gp}
that in neutron-rich spherical nuclei the relative amount of neutrons in comparison to protons 
increases near the surface of the nucleus. This is usually referred to as
neutron-skin effect: In short, the Coulomb barrier that builds up from the positively
charged protons limits the extent that the proton density can stretch out whereas, 
being blind to the Coulomb interaction, the neutrons can be found further away \cite{Horowitz:2000xj}.
The most recent measurement of this phenomenon \cite{Tarbert:2013jze}
for $^{208}$Pb nucleus indicates that the neutron skin does not, unlike its name suggests, have a sharp 
edge but rather a ``halo-like'' character. As shown in the left-hand panel of Figure~\ref{Fig:densityratio},
the proton-to-neutron ratio, $\rho^{{\rm p},A}/\rho^{{\rm n},A}$, does not drop abruptly but the 
fall-off towards the edge of the nucleus is gradual. In this plot (and throughout the rest of this paper) we
have used the parameter values $a_{\rm p}=0.447\,{\rm fm}$, $d_{\rm p}=6.680\,{\rm fm}$ for protons, and $a_{\rm n}=0.55 \pm 0.03\,{\rm fm}$,
$d_{\rm n}=6.70 \pm 0.03 \,{\rm fm}$ for neutrons, taken from Ref.~\cite{Tarbert:2013jze}, and the error band
results from adding the variations caused by the quoted two uncertainties in quadrature.

The Optical Glauber Model \cite{Glauber:1970jm} is a commonly
used tool in heavy-ion collisions \cite{Shukla:2001mb,d'Enterria:2003qs,Miller:2007ri}.
In this approach, the total inelastic cross section $\sigma^{\rm inel}_{AB}(s)$
in collisions of two nuclei $A$ and $B$ with certain center-of-mass energy $\sqrt{s}$
is given by the integral
\begin{equation}
\sigma^{\rm inel}_{AB}({s}) = \int_{-\infty}^{\infty} d^2{\bf b} \left[ 1 - e^{-T_{AB}({\bf b}) \, \sigma^{\rm inel}({s})}\right],
\end{equation}
where ${\bf b}$ is the vector between the centers of the colliding nuclei in transverse plane
(see e.g. Fig.~20 in Ref.~\cite{Helenius:2012wd}), 
$\sigma^{\rm inel}({s})$ is the inelastic nucleon-nucleon cross section\footnote{The total
inelastic cross section should be isospin symmetric as it does not separate the particles
of different charge. That is, we assume
$\sigma^{\rm inel} = \sigma^{\rm inel}_{\rm pp}=\sigma^{\rm inel}_{\rm nn}=\sigma^{\rm inel}_{\rm pn}=\sigma^{\rm inel}_{\rm np}$.},
and $T_{AB}({\bf b})$ is the nuclear overlap function
\begin{eqnarray}
T_{AB}({\bf b}) \equiv \int_{-\infty}^{\infty} d^2{\bf s} \hspace{-0.3cm} & \left[T_A^{\rm p}({\bf s}_1) + T_A^{\rm n}({\bf s}_1) \right] & \\
 & \left[T_B^{\rm p}({\bf s}_2) + T_B^{\rm n}({\bf s}_2) \right] & \hspace{-0.3cm}, \nonumber
\end{eqnarray}
with ${\bf s}_{1,2} \equiv {\bf s} \pm {\bf b}/2$, and
\begin{equation}
T^i_{A}({\bf r}) \equiv \int_{-\infty}^{\infty} dz \rho^{{\rm i},A}({\bf r},z).
\end{equation}
The centrality classes $\mathcal{C}_k$ are defined as ordered impact-parameter intervals
$b_k \leq |{\bf b}| \leq b_{k+1}$, such that a certain percentage $(c_{k+1}-c_k)\%$ of the 
total inelastic cross section accumulates upon integrating,
\begin{eqnarray}
(c_{k+1}-c_k)\% & = & 
\frac{1}{\sigma^{\rm inel}_{AB}} \int_{-\infty}^{\infty} d^2{\bf b} \left[ 1 - e^{-T_{AB}({\bf b}) \, \sigma^{\rm inel}}\right] \nonumber \\
& & \theta\left(b_{k+1} - |{\bf b}|\right) \theta\left(|{\bf b}|-b_k \right).
\end{eqnarray}

Let us now consider a given hard process (e.g. large-$p_T$ direct photon production or equivalent)
whose contribution to $\sigma^{\rm inel}$ is negligible and does thereby not ``interfere'' with our
centrality categorization. 
We write the cross section for such a process within a given centrality class $\mathcal{C}_k$ as
\begin{eqnarray}
d\sigma_{AB}^{\rm hard}({\mathcal{C}_k}) & = &
2\pi \int_{b_k}^{b_{k+1}} dbb \int_{-\infty}^{\infty} d^2{\bf s} \, \sum_{i,j} T_A^{i}({\bf s}_1) T_B^{j}({\bf s}_2) d\sigma_{ij}^{\rm hard}(A,B,{\bf s}_1,{\bf s}_2),
\end{eqnarray}
where the sum runs over different combinations of protons and neutrons. The nucleon-nucleon hard-process
cross sections $d\sigma_{ij}^{\rm hard}(A,B,{\bf s}_1,{\bf s}_2)$ can, in general, depend on the size
of the nuclei (via nuclear shadowing or equivalent \cite{Arneodo:1992wf,Armesto:2006ph,Malace:2014uea}) and 
even on the spatial location of the nucleons inside the nuclei \cite{Vogt:2004hf,Helenius:2012wd}. However, to underscore
the neutron-vs-proton differences alone we assume here that such effects are approximately
constant within each centrality class and write the above cross section as
\begin{eqnarray}
d\sigma_{AB}^{\rm hard}({\mathcal{C}_k}) & = &
\langle nn \rangle_{\mathcal{C}_k} d\sigma_{\rm  nn}^{\rm hard}({\mathcal{C}_k}) + \langle pp \rangle_{\mathcal{C}_k} d\sigma_{\rm pp}^{\rm hard}({\mathcal{C}_k}) \label{eq:hardAA}
 + \langle np \rangle_{\mathcal{C}_k} d\sigma_{\rm np}^{\rm hard}({\mathcal{C}_k}) + \langle pn \rangle_{\mathcal{C}_k} d\sigma_{\rm pn}^{\rm hard}({\mathcal{C}_k}),
\end{eqnarray}
where
\begin{eqnarray}
\langle ij \rangle_{\mathcal{C}_k} \equiv 2\pi \int_{b_k}^{b_k+1} dbb \,  \int_{-\infty}^{\infty} d^2{\bf s} T_{A}^{\rm i}({\bf s}_1) T_{B}^{\rm j}({\bf s}_2),
\label{eq:npAA}
\end{eqnarray}
and $d\sigma_{ij}^{\rm hard}({\mathcal{C}_k})$ refer to nucleon-nucleon cross sections in a given centrality class.
It turns out that for symmetric A+A collisions (for the centrality categories
considered here),
\begin{equation}
\langle np \rangle_{\mathcal{C}_k} = \langle pn \rangle_{\mathcal{C}_k} \approx \sqrt{\langle pp \rangle_{\mathcal{C}_k} \langle nn \rangle_{\mathcal{C}_k} }, 
\end{equation}
and we can visualize a given centrality class $\mathcal{C}_k$ as simply containing events from collisions of two nuclei with
effective number of protons $Z^{AA}_{\rm eff}({\mathcal{C}_k})$ and neutrons $N^{AA}_{\rm eff}({\mathcal{C}_k})$ defined as
\begin{equation}
Z_{\rm eff}^{\rm AA}({\mathcal{C}_k}) \equiv \sqrt{\langle pp \rangle_{\mathcal{C}_k}} \, , \quad
N_{\rm eff}^{\rm AA}({\mathcal{C}_k}) \equiv \sqrt{\langle nn \rangle_{\mathcal{C}_k}} \, . \label{eq:effaa}
\end{equation}
The case of proton-nucleus collisions can be obtained by replacing the 2pF-distribution
for the projectile proton by a delta function $\rho^{p}=\delta^{(3)}({\bf r})$. As a result, the hard-scattering cross section becomes
\begin{equation}
d\sigma^{\rm hard}_{{\rm p}A}({\mathcal{C}_k}) = \langle p \rangle_{\mathcal{C}_k} d\sigma_{\rm pp}^{\rm hard}({\mathcal{C}_k}) + 
                               \langle n \rangle_{\mathcal{C}_k} d\sigma_{\rm pn}^{\rm hard}({\mathcal{C}_k}) \label{eq:hardpA} ,
\end{equation}
with
\begin{eqnarray}
\langle i \rangle_{\mathcal{C}_k} \equiv 2\pi \int_{b_k}^{b_k+1} dbb T_{A}^{\rm i}({\bf b}), 
\end{eqnarray}
and the effective nucleus which the projectile proton ``sees'' consists of
\begin{equation}
Z_{\rm eff}^{\rm pA}({\mathcal{C}_k}) \equiv {\langle p \rangle_{\mathcal{C}_k}} \, , \quad
N_{\rm eff}^{\rm pA}({\mathcal{C}_k}) \equiv {\langle n \rangle_{\mathcal{C}_k}} \, . \label{eq:effpa}
\end{equation}
protons and neutrons, respectively.
To evaluate the effective number of nucleons in each case, we have used 
$\sigma^{\rm inel}(\sqrt{s}=2.76 \, {\rm TeV})=65\,{\rm mb}$ (for Pb+Pb), and
$\sigma^{\rm inel}(\sqrt{s}=5.02 \, {\rm TeV})=70\,{\rm mb}$ (for p+Pb)
for the inelastic nucleon-nucleon cross sections \cite{Antchev:2013iaa}.
The resulting effective proton-to-neutron ratios $Z_{\rm eff}^{\rm pPb}({\mathcal{C}_k})/N_{\rm eff}^{\rm pPb}({\mathcal{C}_k})$
and $Z_{\rm eff}^{\rm PbPb}({\mathcal{C}_k})/N_{\rm eff}^{\rm PbPb}({\mathcal{C}_k})$
computed using Eqs.~(\ref{eq:effaa}) and (\ref{eq:effpa}) in various centrality classes 
are shown in the right-hand panel of Figure~\ref{Fig:densityratio}. While in most central collisions these ratios are
very close to the average value $Z/N=82/126 \approx 0.65$, in very peripheral
bins the relative amount of neutrons grows.
Since the edges of the nuclei are always inside the integration domain in Eq.~(\ref{eq:npAA}),
the effect of neutron skin in lead-lead starts to be visible in more central collision than 
in the case of proton-lead.
This observation also hints that our assumption of a point-like proton makes the centrality-dependent
effects slightly weaker than what would be obtained by assigning the proton with a finite size.

\section{Effects of Neutron Skin in $W^\pm$ Production}

To make the variations in proton-to-neutron ratio visible, an observable for which
$d\sigma_{\rm nn}^{\rm hard} \neq d\sigma_{\rm pp}^{\rm hard} \neq d\sigma_{\rm pn}^{\rm hard}$
is required. As the only significant difference\footnote{We neglect the photon distribution \cite{Martin:2004dh,Ball:2013hta}}
between the protons and neutrons is their
$u$ and $d$ valence-quark content, we need a probe which couples differently
to $u$ and $d$ flavours. Here, we will
consider the production of inclusive charged leptons $\ell^\pm$ from $W^\pm \rightarrow \ell^\pm \nu$ decays.
This process is theoretically particularly well understood, the collinearly factorized
perturbative QCD calculations known up to next-to-next-to-leading order \cite{Anastasiou:2003ds,Catani:2009sm}, and
the state-of-the-art calculations incorporate also next-to-leading order electroweak corrections \cite{Li:2012wna}
on top of this. The existing minimum bias LHC measurements for this process in proton-lead and lead-lead collisions
are roughly consistent with the pQCD predictions \cite{Chatrchyan:2012nt,ATLASPbPb,CMS:2014kla}.

In the narrow-width approximation, accurate in the asymptotic limit when the decay width $\Gamma_{\rm W}$ of the ${\rm W}^\pm$ boson is much less than its mass $M_{\rm W}$, 
the leading-order expressions \cite{Aurenche:1980tp,Baer:1990qy} for the charged-lepton rapidity ($y$) and transverse
momentum ($p_T$) distribution can be cast as
\begin{eqnarray}
\frac{d\sigma^{\ell^\mp}}{dydp_T} & \approx & \frac{\pi^2}{24s} \left( \frac{\alpha_{\rm em}}{\sin^2\theta_{\rm W}} \right)^2 
\frac{1}{M_{\rm W} \Gamma_{\rm W}}
\frac{p_T}{\sqrt{1-4p_T^2/M_{\rm W}^2}} \sum _{i,j} |V_{ij}|^2 \, \delta \left( e_{q_i} + e_{\overline{q}_j} \pm 1 \right) \\
& \times & 
\left\{
\alpha^\pm \left[ f_{q_i}^{\rm A}(\xi_1^+,Q^2) f_{\overline{q}_j}^{\rm B}(\xi_2^+,Q^2) + f_{\overline{q}_j}^{\rm A}(\xi_1^-,Q^2) f_{q_i}^{\rm B}(\xi_2^-,Q^2) \right] \right. \nonumber \\
& + & \left. \,\,\, \alpha^\mp \left[ f_{q_i}^{\rm A}(\xi_1^-,Q^2) f_{\overline{q}_j}^{\rm B}(\xi_2^-,Q^2) + f_{\overline{q}_j}^{\rm A}(\xi_1^+,Q^2) f_{q_i}^{\rm B}(\xi_2^+,Q^2)  \right]
\right\} \nonumber
\end{eqnarray}
where $\alpha^\pm  =  1 \pm (1-4p_T^2/M_{\rm W}^2)^{1/2}$, 
and the symbols $\alpha_{\rm em}$, $\theta_{\rm W}$ and $V_{ij}$ denote the fine-structure constant, weak-mixing angle
and Cabibbo-Kobayashi-Maskawa matrix, respectively.
The sum over all partonic flavors is restricted by the $\delta$ function which selects only those
combinations of quarks and antiquarks for which the electric charges $e_{q_i}$, $e_{\overline{q}_j}$ sum up 
correctly to the charge of the lepton. The momentum arguments of the PDFs 
for quarks $f_{q_i}^{\rm A}(x,Q^2)$ and anti quarks $f_{\overline{q}_i}^{\rm A}(x,Q^2)$ are given by
\begin{equation}
 \xi_1^\pm \equiv \frac{M_{\rm W}^2 e^y \,\,\,\, }{2p_T\sqrt{s}} \left[1 \mp \sqrt{1-4p_T^2/M_{\rm W}^2}\right],
 \quad
 \xi_2^\pm \equiv \frac{M_{\rm W}^2 e^{-y}}{2p_T\sqrt{s}} \left[1 \pm \sqrt{1-4p_T^2/M_{\rm W}^2}\right].
 \label{eq:xs}
\end{equation}
To account for the centrality-dependence of the hard-scattering cross sections,
we use PDFs $f^{\rm Pb,{\mathcal{C}_k}}_i(x,Q^2)$ for the lead nucleus defined as
\begin{equation}
f_i^{\rm Pb,{\mathcal{C}_k}}(x,Q^2) \equiv Z_{\rm eff}^{{\rm pPb,PbPb}}({\mathcal{C}_k}) f_i^{\rm p,{\mathcal{C}_k}}(x,Q^2) + N_{\rm eff}^{{\rm pPb,PbPb}}({\mathcal{C}_k}) f_i^{\rm n,{\mathcal{C}_k}}(x,Q^2),
\end{equation}
where $f_i^{\rm p,{\mathcal{C}_k}}$ and $f_i^{\rm n,{\mathcal{C}_k}}$ are proton and neutron PDFs
in a given centrality class ${\mathcal{C}_k}$, the latter obtained from the former based on the
isospin symmetry (e.g. $f_u^{\rm n,{\mathcal{C}_k}}=f_d^{\rm p,{\mathcal{C}_k}}$). Other nuclear
effects like shadowing can be incorporated 
as multiplicative correction factors $R_i^{{\mathcal{C}_k}}(\xi,Q^2)$ on the proton PDFs,
\begin{equation}
f_i^{\rm p,{\mathcal{C}_k}}(\xi,Q^2) = R_i^{{\mathcal{C}_k}}(\xi,Q^2) f_i^{\rm p}(\xi,Q^2).
\end{equation}
In the case of minimum bias collisions these correction factors and their uncertainties have
been estimated in several global fits \cite{Hirai:2007sx,Eskola:2009uj,deFlorian:2011fp,Kusina:2014wwa},
and models for their possible centrality dependence exist \cite{Vogt:2004hf,Helenius:2012wd}.
However, in this letter, we will consider only ratios $d\sigma(\ell^+)/d\sigma(\ell^-)$ and
nuclear modifications like this are expected to largely cancel even if they were centrality
dependent: At a high factorization scale like $Q^2 \propto M_{\rm W}^2$ involved here, most of
the sea quarks originate from collinear gluon splittings ($g\rightarrow q\overline{q}$) which is
a flavor-independent process for light quarks.
Furthermore, the mutually very similar nuclear effects observed in charged-lepton \cite{Arneodo:1992wf,Malace:2014uea} 
and neutrino deep-inelastic scattering \cite{Paukkunen:2013grz} on Pb nucleus 
indicate that the nuclear corrections for the valence quarks are also
approximately equal, $R^{{\mathcal{C}_k}}_{u_{\rm v}}(\xi,Q^2) \approx R^{{\mathcal{C}_k}}_{d_{\rm v}}(\xi,Q^2)$. 
As a result, the overall nuclear corrections for $\ell^+$ and $\ell^-$ production must be 
mutually very alike and largely cancel upon taking the ratio, separately in each centrality class.
In fact, the results shown in plots below have been obtained by setting $f_i^{\rm p,{\mathcal{C}_k}} = f_i^{\rm p}$
where $f_i^{\rm p}$ are free proton PDFs for which we have used the general-purpose 
\texttt{CT10NLO} parametrization \cite{Lai:2010vv}. 
Another attractive feature that the ratio $d\sigma(\ell^+)/d\sigma(\ell^-)$ entails is that 
many experimental systematic uncertainties can be expected to cancel out. Furthermore, one does not
need an absolute normalization which would need further Glauber modeling.

The results that follow have been obtained by using the \texttt{MCFM} Monte-Carlo code \cite{Campbell:2010ff,MCFM} 
at next-to-leading order accuracy with the factorization and renormalization scales $Q^2$ fixed to $M_{\rm W}$.
To mimic a realistic experimental situation, we integrate over the charged lepton transverse momentum with $p_T>25\,{\rm GeV}$. 

\subsection{Proton-Lead Collisions}

\begin{figure}[ht!]
\center
\includegraphics[width=0.52\textwidth]{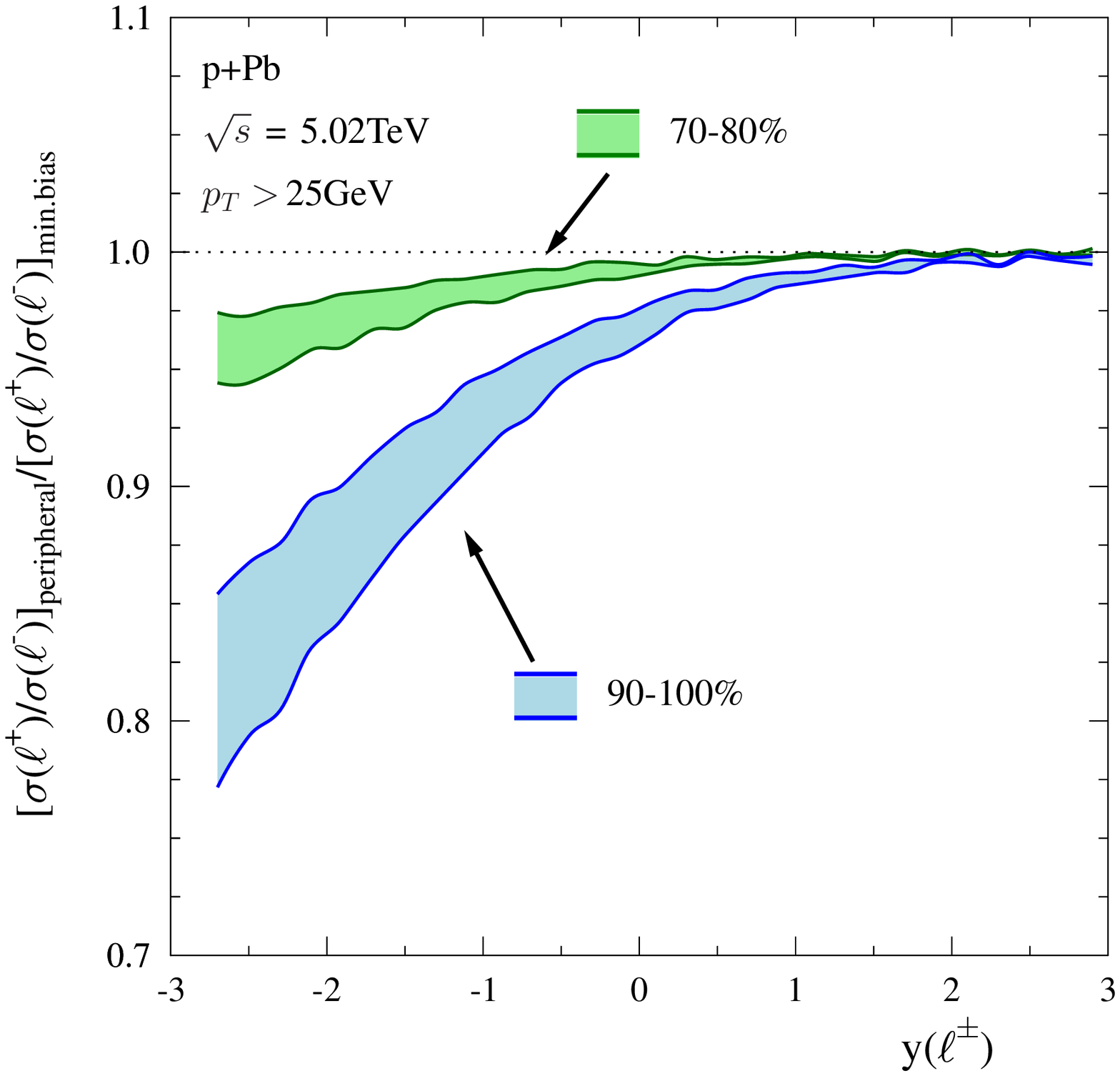}
\hspace{-1.0cm}
\includegraphics[width=0.52\textwidth]{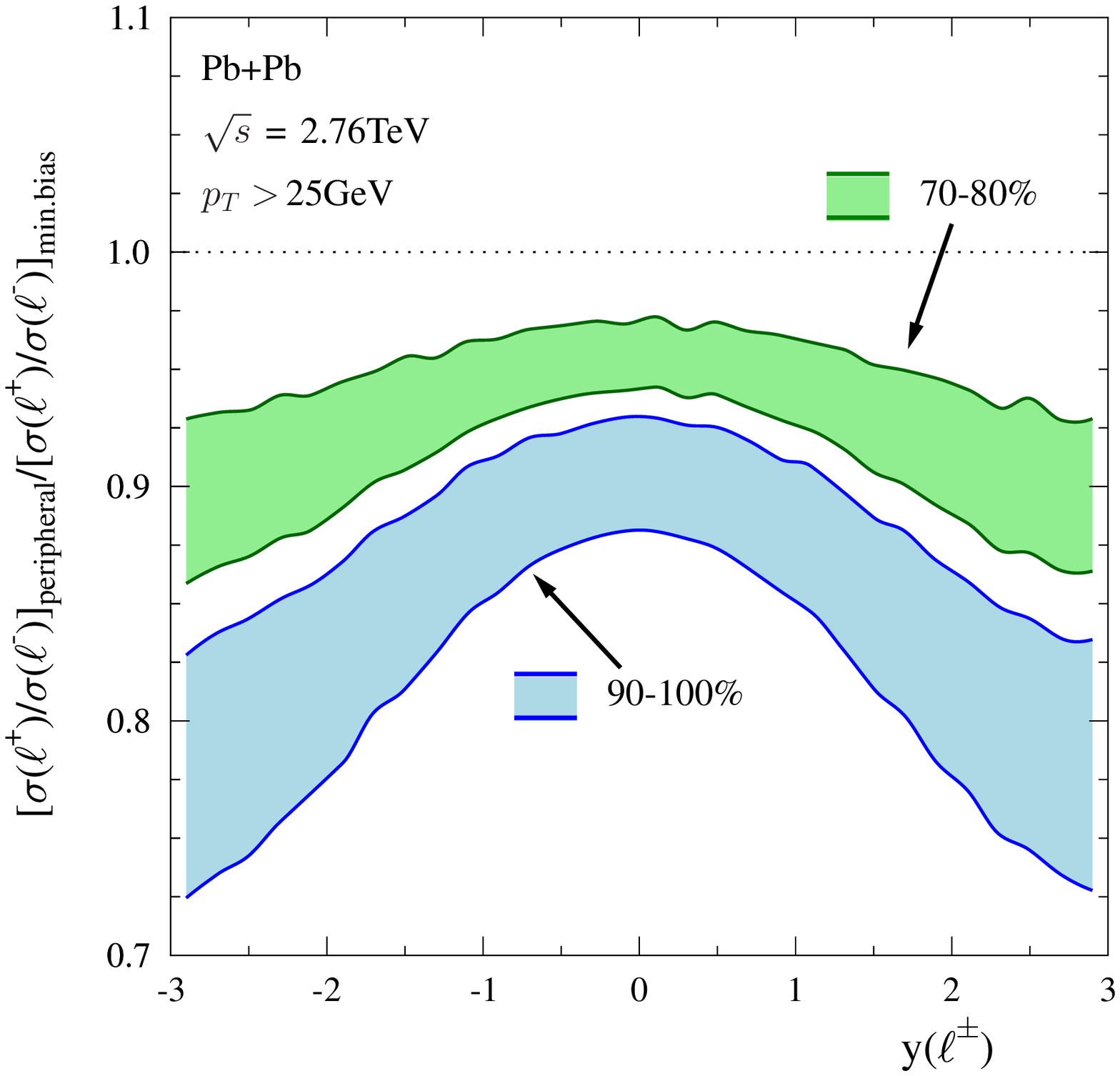}
\caption{Left-hand panel: The ratio $d\sigma^{\ell^+}/d\sigma^{\ell^-}$ in proton-lead collisions for
two peripheral centrality class normalized by the corresponding ratio
in the minimum bias collisions. Right-hand panel: As the left-hand panel, but
for lead-lead collisions. The small wiggles are residual fluctuations from the Monte-Carlo integrations
in \texttt{MCFM}.
}
\label{Fig:pPb_central_to_pheri}
\end{figure}

The effect of the neutron skin will be most pronounced in the kinematic region where 
the large-$x$ nuclear valence quarks $f_{u_{\rm v}}^{\rm p,{\mathcal{C}_k}}(x)$ and $f_{d_{\rm v}}^{\rm p,{\mathcal{C}_k}}(x)$
are of importance. From Eq.~(\ref{eq:xs}) we see that this happens
towards negative values of $y$ (the ``backward'' direction).
The resulting centrality dependence of ratios $d\sigma(\ell^+)/d\sigma(\ell^-)$ in p+Pb collisions are
illustrated in the left-hand panel of Figure~\ref{Fig:pPb_central_to_pheri} by comparing two peripheral
classes 70-80\% and 90-100\% to the minimum bias one, 0-100\%. Since the main contributions to the 
cross sections come from the $u\overline{d}$ and $d\overline{u}$ partonic channels \cite{Kusina:2012vh}
and the sea-quark distributions at small $x$ and large factorization scale $Q^2=M_{\rm W}^2$ are approximately
flavor independent, $f_{\overline{d}}^{\rm p}(x,Q^2) \approx f_{\overline{u}}^{\rm p}(x,Q^2)$, we can approximate
\begin{eqnarray}
 \hspace{-0.5cm}
 \frac{d\sigma_{\rm p+Pb}^{\ell^+}}{d\sigma_{\rm p+Pb}^{\ell^-}}\Big|_{{y \ll 0}} & \approx & 
       \label{eq:backpPb} 
\left(\frac{\alpha^-}{\alpha^+}\right)
 \frac{\left[ Z_{\rm eff}^{{\rm pPb}}({\mathcal{C}_k})/N_{\rm eff}^{{\rm pPb}}({\mathcal{C}_k}) \right] f_{u_{\rm v}}^{\rm p,{\mathcal{C}_k}}(\xi_2^-,Q^2) + f_{d_{\rm v}}^{\rm p,{\mathcal{C}_k}}(\xi_2^-,Q^2)}
      {\left[ Z_{\rm eff}^{{\rm pPb}}({\mathcal{C}_k})/N_{\rm eff}^{{\rm pPb}}({\mathcal{C}_k}) \right] f_{d_{\rm v}}^{\rm p,{\mathcal{C}_k}}(\xi_2^-,Q^2) + f_{u_{\rm v}}^{\rm p,{\mathcal{C}_k}}(\xi_2^-,Q^2)}. 
\end{eqnarray}
As $f_{d_{\rm v}}^{\rm p,{\mathcal{C}_k}}(x,Q^2)/f_{u_{\rm v}}^{\rm p,{\mathcal{C}_k}}(x,Q^2)<1$,
the derivative of this expression with respect to $Z_{\rm eff}^{{\rm pPb}}({\mathcal{C}_k})/N_{\rm eff}^{{\rm pPb}}({\mathcal{C}_k})$
is positive and, in line with Figure~\ref{Fig:pPb_central_to_pheri}, the ratio $d\sigma^{\ell^+}/d\sigma^{\ell^-}$
decreases towards more peripheral collisions (since $Z_{\rm eff}^{{\rm pPb}}({\mathcal{C}_k})/N_{\rm eff}^{{\rm pPb}}({\mathcal{C}_k})$ decreases).
Towards positive values of $y$ (the ``forward'' direction), we have 
\begin{eqnarray}
 \frac{d\sigma_{\rm p+Pb}^{\ell^+}}{d\sigma_{\rm p+Pb}^{\ell^-}}\Big|_{{y \gg 0}} \hspace{-0.2cm} & \approx & 
 \left(\frac{\alpha^-}{\alpha^+}\right) \frac{f_{u_{\rm v}}^{\rm p}(\xi_1^+,Q^2)}{f_{d_{\rm v}}^{\rm p}(\xi_1^+,Q^2)}, \label{eq:forpPb}
 \end{eqnarray}
where we have assumed that the small-$x$ sea quark distributions in Pb nucleus are approximately flavor independent
at large $Q^2$, $f_{\overline u}^{\rm Pb,{\mathcal{C}_k}}(x,Q^2) \approx f_{\overline d}^{\rm Pb,{\mathcal{C}_k}}(x,Q^2)$. The independence of $Z_{\rm eff}^{{\rm pPb}}({\mathcal{C}_k})/N_{\rm eff}^{{\rm pPb}}({\mathcal{C}_k})$
explains why the centrality dependence of $d\sigma^{\ell^+}/d\sigma^{\ell^-}$ virtually disappears
towards large values of $y$. Currently, no centrality classified proton-lead data for $W^\pm$ production
are available.

\subsection{Lead-Lead Collisions}

The right-hand panel of Figure~\ref{Fig:pPb_central_to_pheri} presents the results in the case of symmetric Pb+Pb collisions.
As earlier, sufficiently far away from the midrapidity, $|y|\gg0$, we can approximate
 \begin{eqnarray}
 \frac{d\sigma_{\rm Pb+Pb}^{\ell^+}}{d\sigma_{\rm Pb+Pb}^{\ell^-}}\Big|_{{|y| \gg 0}} & \approx & 
\left(\frac{\alpha^-}{\alpha^+}\right)
 \frac{\left[ Z_{\rm eff}^{{\rm PbPb}}({\mathcal{C}_k})/N_{\rm eff}^{{\rm PbPb}}({\mathcal{C}_k}) \right]f_{u_{\rm v}}^{\rm p,{\mathcal{C}_k}}(x,Q^2) + f_{d_{\rm v}}^{\rm p,{\mathcal{C}_k}}(x,Q^2)}
      {\left[ Z_{\rm eff}^{{\rm PbPb}}({\mathcal{C}_k})/N_{\rm eff}^{{\rm PbPb}}({\mathcal{C}_k}) \right]f_{d_{\rm v}}^{\rm p,{\mathcal{C}_k}}(x,Q^2) + f_{u_{\rm v}}^{\rm p,{\mathcal{C}_k}}(x,Q^2)} 
       \label{eq:backPbPb}
 \end{eqnarray}
where now $x=\xi_2^-$ for $y\ll 0$, and $x=\xi_1^+$ for $y \gg 0$. Similarly as in the case 
of proton-lead, the dependence of $Z_{\rm eff}^{{\rm PbPb}}({\mathcal{C}_k})/N_{\rm eff}^{{\rm PbPb}}({\mathcal{C}_k})$
explains the stronger suppression of $d\sigma(\ell^+)/d\sigma(\ell^-)$ towards peripheral collisions. 
In comparison to the proton-lead collisions, the effect is better visible even at $y=0$ since both the sea-valence
and valence-sea scatterings depend on $Z_{\rm eff}^{{\rm PbPb}}({\mathcal{C}_k})/N_{\rm eff}^{{\rm PbPb}}({\mathcal{C}_k})$
while in proton-lead collisions this happens only for sea-valence contribution.

\begin{figure}[ht!]
\center
\includegraphics[width=0.60\textwidth]{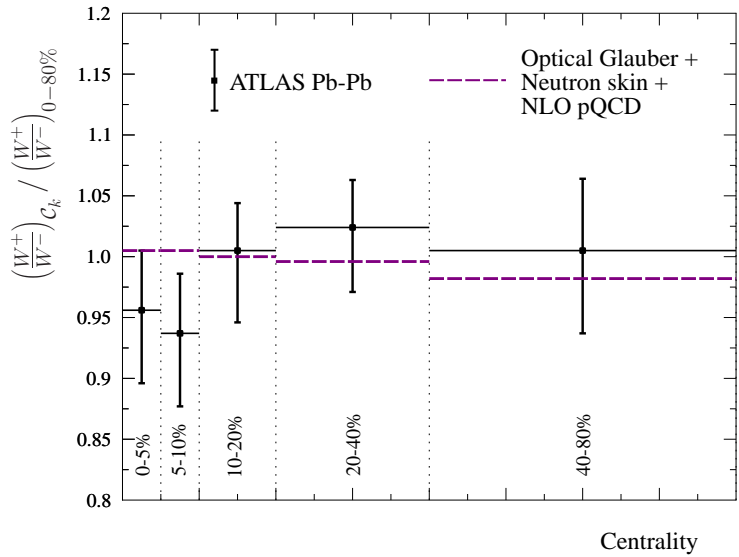}
\caption{The centrality dependence of $W^+/W^-$ ratio as measured by
the ATLAS collaboration \cite{ATLASPbPb} compared to the calculation presented in this paper.
The reported experimental values of $W^+/W^-$ ratio in each centrality class have been
normalized to the average one $(W^+/W^-)_{0-80\%}=1.03$ and the uncertainties 
have been obtained by adding in quadrature the statistical and systematic uncertainties.
}
\label{Fig:ComparisonATLAS}
\end{figure}

Currently, the most accurate experimental measurements of $W^\pm$ production in lead-lead
collisions come from the ATLAS collaboration \cite{ATLASPbPb}. The data for $W^+/W^-$ ratio
are plotted in Figure~\ref{Fig:ComparisonATLAS} and compared to the calculation described
in this letter. The data, as well as the \texttt{MCFM} computation, include all $W^\pm$ events
within the kinematic region restricted by the lepton transverse momentum $p_T > 25 \, {\rm GeV}$,
pseudorapidity interval $|\eta^{\ell^\pm}|<2.5$, missing transverse momentum $p_T^{\rm missing} > 25 \, {\rm GeV}$, 
and the transverse mass of the $\ell^\pm \nu$ system $m_T > 40 \, {\rm GeV}$.
While our calculation is consistent with the data, $\chi^2/N_{\rm data} \approx 0.6$ ($N_{\rm data}$ is
the number of data points), the measurements are clearly not accurate enough and have a too coarse centrality categorization
to draw conclusions to any direction at this stage.

\section{Summary and Outlook}

We have discussed the generic effects that mutually different spatial distributions
of protons and neutrons in heavy nuclei are expected to induce in production of W$^\pm$ bosons
at the LHC. The proton density is known to fall off more rapidly than that of neutrons
towards the surface of neutron-rich nuclei like $^{208}$Pb which, as we have demonstrated,
correlates with the sign of the produced W$^\pm$ boson. Thus, the W$^\pm$ production could be used
to benchmark different experimental definitions of centrality at the LHC.
We stress that in this paper we have considered the centrality from a purely geometric viewpoint
neglecting e.g. possible smearing of the effect that could be caused by event-by-event fluctuations.
Therefore, our results should be taken as first, rough estimates of the expected
systematics if a given experimental centrality-selection method truly reflects the collision geometry.

The effects caused by neutron skin are, of course, not limited to W$^\pm$ production but 
related phenomena are to be expected e.g. in the case of high-$p_T$ photons
and charged hadrons which can also be
measured at lower center-of-mass energies like those available at the Relativistic Heavy-Ion
Collider (RHIC). In the case of lepton-nucleus deeply-inelastic scattering, the
neutron skin should affect differently the neutral- and charged-current reactions.
Thus, the neutron-skin effect could serve as a handle to study the centrality also
at planned deeply-inelastic scattering experiments like the Electron-Ion Collider \cite{Aschenauer:2014cki}
or LHeC \cite{AbelleiraFernandez:2012cc}.

\section*{Acknowledgments}

I thank Markus Kortelainen, Sami R\"as\"anen, Kari~J.~Eskola, Tuomas Lappi and Harri Niemi for discussions.

\end{document}